# A hierarchy of compatibility and comeasurability levels in quantum logics with unique conditional probabilities

Gerd Niestegge

Zillertalstrasse 39, 81373 Muenchen, Germany
gerd.niestegge@web.de

*Abstract*: In the quantum mechanical Hilbert space formalism, the probabilistic interpretation is a later ad-hoc add-on, more or less enforced by the experimental evidence, but not motivated by the mathematical model itself. A model involving a clear probabilistic interpretation from the very beginning is provided by the quantum logics with unique conditional probabilities. It includes the projection lattices in von Neumann algebras and here probability conditionalization becomes identical with the state transition of the Lüders - von Neumann measurement process. This motivates the definition of a hierarchy of five compatibility and comeasurability levels in the abstract setting of the quantum logics with unique conditional probabilities. Their meanings are: the absence of quantum interference or influence, the existence of a joint distribution, simultaneous measurability, and the independence of the final state after two successive measurements from the sequential order of these two measurements. A further level means that two elements of the quantum logic (events) belong to the same Boolean subalgebra. In the general case, the five compatibility and comeasurability levels appear to differ, but they all coincide in the common Hilbert space formalism of quantum mechanics, in von Neumann algebras, and in some other cases.

*Key words*: Quantum measurement, conditional probability, quantum logic, operator algebras.

*PACS*: 03.65.Ta, 03.67.-a

## 1 Introduction

For a better understanding of quantum mechanics and particularly of its measurement process, the study of more general theories may be useful and help to identify typical properties distinguishing quantum mechanics or to reconstruct the mathematical formalism of quantum mechanics from a few basic principles or axioms. A most general and abstract framework for the study of the quantum measurement process are the quantum logics with unique conditional probabilities introduced in [1]. The state transition of the Lüders - von Neumann measurement process is a special case of the probability conditionalization in this general setting, motivating the definition of five levels of compatibility and comeasurability with the following meanings for a pair of quantum events, i.e., a pair of elements *e* and *f* in the quantum logic:

(1) There is no influence of *e* on *f* or, equivalently, the joint distribution of (*e*,*f*) exists for all states, assuming that *e* is measured first.
(2) Compatibility: There is neither an influence of *e* on *f* nor an influence of *f* on *e* or, in other words, these two events do not involve any quantum interference or, equivalently, the joint distributions of (*e*,*f*) as well as of (*f*,*e*) both always exist.
(3) Weak comeasurability: The two events *e* and *f* are simultaneously measurable. I.e., the joint distributions of (*e*,*f*) and (*f*,*e*) always exist and the one of (*e*,*f*) is identical with the one of (*f*,*e*) up to the exchange of the variables.
(4) Strong comeasurability: The final states after two successive measurements of *e* and *f* are identical in the two cases when *e* is measured first and *f* second or when *f* is measured first and *e* second.
(5) Algebraic compatibility: The two events *e* and *f* belong to a Boolean subalgebra of the quantum logic.

More precise definitions of these compatibility and comeasurability levels will be presented in sections 4 to 8, and it will be seen that there is a certain hierarchy among them in the general case; the fifth one (algebraic compatibility) is the strongest level implying all the other ones and the first one is the weakest level. In Hilbert space quantum mechanics, all five compatibility and comeasurability levels are equivalent and identical with the common concept of operator commutation.

Moreover, the second and the third level (compatibility and weak comeasurability) coincide in the general case if the conditional probabilities satisfy a certain symmetry condition which was originally discovered by Alfsen and Shultz [2] from purely mathematical reasons. They used it to derive a Jordan product in their non-commutative spectral theory and, in this way, the condition appears in some axiomatic approaches to quantum mechanics [3], [4]. The same condition is studied in [5] where it is shown that it implies the absence of third-order interference as introduced by Sorkin [6].

Furthermore, the relation between comeasurability and a potential logical and-operation on the quantum logic is considered, and a setting which is more general than Hilbert space quantum mechanics, but in which the five compatibility and comeasurability levels still coincide is presented.

The paper starts with a brief survey of the quantum logics with unique conditional probabilities (sections 2 and 3). The different levels of compatibility and comeasurability are introduced and studied in the subsequent sections beginning with the weakest one and then moving on to the stronger ones. In section 4, compatibility and a weaker asymmetric version of it are defined by the absence of quantum interference or influence. In section 5, the joint distributions existing in the case of compatibility are considered, providing the motivation for the weak comeasurability which is defined in section 6 and means simultaneous measurability. It becomes equivalent to the compatibility of section 4 in the case of the validity of the symmetry condition for the conditional probabilities. The strong comeasurability is introduced in section 7; its meaning is the independence of the final state after two successive measurements from the sequential order of these two measurements. The algebraic compatibility considered in section 8 represents the strongest one among these levels of compatibility and comeasurability. Their relation to a potential logical and-operation for pairs of events in the quantum logic is studied in section 9. The coincidence of all five compatibility and comeasurability levels in Hilbert space quantum mechanics is shown in section 10 and, in section 11, a more general setting where they still coincide is presented. This last section is the only one which exceeds the basic layer of the previous ones a little.

## 2  The quantum logic

A quantum logic is the mathematical model of a system of quantum events or propositions. Logical approaches use the name "proposition", while the name "event" is used in probability theory and will also be preferred in the present paper. The concrete quantum logic of standard quantum mechanics is the system of closed linear subspaces of a Hilbert space or, more generally, the projection lattice in a von Neumann algebra.

Usually, an abstract quantum logic is assumed to be an orthomodular partially ordered set and, very often, it is also assumed that it is lattice. For the purpose of the present paper, however, a more general and simpler mathematical structure without order relation is sufficient. Only an orthocomplementation, an orthogonality relation and a sum operation defined for orthogonal events are needed. The orthocomplementation represents the logical negation, orthogonality means mutual exclusivity, and the sum represents the logical and-operation in the case of mutual exclusivity. The precise axioms were presented in [1] and look as follows.

The quantum logic $E$ is a set with distinguished elements $0$ and $\mathbb{1}$, an orthogonality relation $\perp$ and a partial binary operation $+$ such that the following axioms hold for $e,f,g \in E$:

(OS1)  *If $e \perp f$, then $f \perp e$; i.e., the relation $\perp$ is symmetric.*
(OS2)  *$e+f$ is defined for $e \perp f$, and then $e+f=f+e$; i.e., the sum operation is commutative.*
(OS3)  *If $g \perp e$, $g \perp f$, and $e \perp f$, then $g \perp e+f$, $f \perp g+e$ and $g+(e+f)=(g+e)+f$; i.e., the sum operation is associative.*
(OS4)  *$0 \perp e$ and $e+0=e$ for all $e \in E$.*
(OS5)  *For every $e \in E$, there exists a unique $e' \in E$ such that $e \perp e'$ and $e+e'=\mathbb{1}$.*
(OS6)  *There exists $d \in E$ such that $e \perp d$ and $e+d=f$ if and only if $e \perp f\,'$.*

Then $0'=\mathbb{1}$ and $e''=e$ for $e \in E$. Note that an orthomodular partially ordered set satisfies these axioms with the two definitions

  (i)   *$e \perp f$ iff $f \leq e'$*
  (ii)  *The sum $e+f$ is the supremum of $e$ and $f$ for $e \perp f$.*

The supremum exists in this case due to the orthomodularity. In particular, a Boolean lattice or Boolean algebra satisfies all the above axioms.

However, in the general case, $f \perp e_k$ implies $f \perp \sum e_k$, but $f \perp \sum e_k$ does not imply that $f \perp e_k$ for each $k$, where $e_1,e_2,...e_n$ are $n$ pairwise orthogonal events and $f$ is a further event.

## 3  Conditional probabilities

A state is a map $\mu: E \to [0,1]$ such that $\mu(\mathbb{1})=1$ and $\mu(e+f) = \mu(e) + \mu(f)$ for orthogonal pairs $e$ and $f$ in $E$. Then $\mu(0)=0$ and $\mu(e_1+...+e_k) = \mu(e_1)+...+\mu(e_k)$ for pairwise orthogonal elements $e_1,...,e_k$ in $E$. Denote by $S$ the set of all states on $E$. With a state $\mu$ and $\mu(e)>0$ for an $e \in E$, another state $\nu$ is called a conditional probability of $\mu$ under $e$ if $\nu(f) = \mu(f)/\mu(e)$ holds for all $f \in E$ with $f \perp e'$. Furthermore, the following axioms were introduced in [1].

(UC1)  *If $e,f \in E$ and $\mu(e)=\mu(f)$ for all $\mu \in S$, then $e=f$.*
(UC2)  *If $e \in E$ and $\mu \in S$ with $\mu(e)>0$, there is one and only one conditional probability of $\mu$ under $e$.*

If these axioms are satisfied, *E* is called a *UCP space* – named after the major feature of this mathematical structure which is the existence of the <u>u</u>nique <u>c</u>onditional <u>p</u>robability – and the elements in *E* are called events. The unique conditional probability of μ under *e* is denoted by $\mu_e$ and, in analogy with probability theory, μ(*f*|*e*) is often written instead of $\mu_e$(*f*) with *f*∈ *E*. The above two axioms imply that there is a state μ∈ *S* with μ(*e*)=1 for each event *e*≠0, that the difference *d* in (OS6) becomes unique, and that *e*⊥*e* iff *e*⊥$\mathbb{1}$ iff *e*=0 (*e*∈ *E*).

In the remaining part of the present paper it will always be assumed that *E* is a UCP space. Although $\mu_e$ and μ(*f*|*e*) are defined only for μ(*e*)>0, the restriction μ(*e*)>0 will not be made when the products μ(*e*)$\mu_e$ or μ(*e*)μ(*f*|*e*) are considered; these products shall be defined as μ(*e*)$\mu_e$:=0 and μ(*e*)μ(*f*|*e*):=0 for μ(*e*)=0.

A typical example of a UCP space is the projection lattice *E* in a von Neumann algebra *M* without type $I_2$ part; *E* = {*e*∈ *M*: *e*\*=*e*=$e^2$}. The conditional probability then has the shape

$$\mu_e(f) = \mu(f|e) = \frac{1}{\mu(e)} \hat{\mu}(efe) \tag{1}$$

with *e*,*f*∈ *E*, μ∈ *S* and μ(*e*)>0. Note that $\hat{\mu}$ on *M* is the unique positive linear extension of the state μ originally defined only on the projection lattice; this extension exists by Gleason's theorem [7] and its later enhancements to finitely additive states and arbitrary von Neumann algebras [8], [9], [10], [11]. The linear extension $\hat{\mu}$ does not exist if *M* contains a type $I_2$ part.

For the proof of equation (1), suppose that the state ν on *E* is a version of the conditional probability of the state μ under *e* and use the identity *f*=*efe*+*efe'*+*e'fe*+*e'fe'*, where *e'*=$\mathbb{1}$-*e*. From ν(*e'*)=0 and the Cauchy-Schwarz inequalty applied with the positive linear functional $\hat{\nu}$ it follows that 0 = $\hat{\nu}(efe')$ = $\hat{\nu}(e'fe)$ = $\hat{\nu}(e'fe')$ such that ν(*f*) = $\hat{\nu}(efe)$. By the spectral theorem, *efe* can be approximated (in the norm topology) by linear combinations of elements in {*d*∈ *E*:*d*⊥*e'*} = {*d*∈ *E*:*d*≤*e*} for which ν coincides with μ/μ(*e*). The continuity of $\hat{\nu}$ (due its positivity) then implies ν(*f*) = $\hat{\nu}(efe)$ = $\hat{\mu}(efe)/\mu(e)$. Therefore the conditional probability must have this shape and its uniqueness is proved. Its existence follows from *efe*≥0 and *efe*=*f* for *f*≤*e*, since then ν(*f*) := $\hat{\mu}(efe)/\mu(e)$ indeed owns all the properties of the conditional probability.

Equation (1) reveals the link to the Lüders – von Neumann quantum measurement process. The transition from a state μ to the conditional probability $\mu_e$ is identical with the transition from the state prior to the measurement to the state after the measurement where *e* represents the measurement result. This motivates most of the definitions of the compatibility and comeasurability levels in the following sections.

## 4  Compatibility

A quantum measurement with the outcome *e* transforms the initial state μ to the new state $\mu_e$ which is the conditional probability under *e*. This transformation assumes that the measurement is executed and that the observer knows the result of the measurement. A different situation occurs when the measurement is executed, but the result in not known (yet); the observer may just not (yet) have looked at the pointer of the measurement apparatus. Then the initial state μ is transformed to the state μ(*e*)$\mu_e$+μ(*e'*)$\mu_{e'}$ assuming that the measurement is a simple test of *e* versus *e'*.

With classical conditional probabilities, this state is identical with the initial state μ. However, considering again the von Neuman algebras, this becomes the state which maps $f$ to $\hat{\mu}(efe) + \hat{\mu}(e'fe')$ and which is not identical with the the initial state in many cases. The difference between this state and the initial state μ is a very important quantum phenomenon distinguishing quantum mechanics from the classical case. It is responsible for the well-known interference effects in quantum mechanics. If μ($f$) is not identical with μ($f|e$)μ($e$) + μ($f|e'$)μ($e'$), there is some kind of interference present in the state μ between the events $e$ and $f$.

Kläy, Randall and Foulis [12] introduced another wording for this situation saying that the state μ exhibits an influence of the event $e$ on the event $f$. They considered the more specific situation of a compound system where $e$ and $f$ belong to different subsystems. Actually the use of the word "influence" goes back to Dirac. That the meaning of "influence" is very special here becomes clear when considering that this type of influence cannot occur with classical probabilities as mentioned above already. The presence of an influence means that the execution of a simple test of $e$ versus $e'$ has an effect on the probability at which $f$ occurs even when the actual test outcome ($e$ or $e'$) is unknown. Note that the classical concept of stochastic dependence describes a totally different kind of influence between events.

The following notation is now introduced: $e \rightarrow_\mu f$ shall mean that the identity μ($f$) = μ($f|e$)μ($e$) + μ($f|e'$)μ($e'$) holds for the state μ and the events $e$ and $f$. If $e \rightarrow_\mu f$ as well as $f \rightarrow_\mu e$ hold, $e \leftrightarrow_\mu f$ is written. This symmetric property shall be called relative compatibility of the two events $e$ and $f$ in the state μ. In this case, there is no interference or no influence in the above sense present between the events $e$ and $f$ in the state μ; the state μ exhibits neither an influence of $e$ on $f$ nor an influence of $f$ on $e$.

Moreover, $e \rightarrow f$ shall mean that $e \rightarrow_\mu f$ holds for all states μ, and $e \leftrightarrow f$ shall mean that $e \leftrightarrow_\mu f$ holds for all states μ. In the latter case, the events $e$ and $f$ shall be called compatible.

## 5  Joint distributions

The conditional probabilities give rise to a definition of a joint distribution in a quite natural way without having to assume that $E$ is a lattice. The product μ($e$)μ($f|e$) is the probability that the outcome of a first measurement is $e$ and that the outcome of a second subsequent measurement is $f$, when μ is the initial state of the system under consideration before the two measurements. Therefore, a classical probability distribution $p$ on the classical two-bit space {0,1}×{0,1} shall be called the joint distribution of the event pair ($e,f$) in the state μ if the following six identities hold:

$$p(1,1) = \mu(e)\mu(f|e), \qquad p(1,0) = \mu(e)\mu(f'|e),$$
$$p(0,1) = \mu(e')\mu(f|e'), \qquad p(0,0) = \mu(e')\mu(f'|e'),$$

$$p(1,0) + p(1,1) = \mu(e), \qquad p(0,1) + p(1,1) = \mu(f).$$

The function $p$ represents the joint probability distribution over all the possible outcomes of two successive measurements where the first one tests $e$ versus $e'$ and the second one $f$ versus $f'$. Note that the joint distribution of the event pair ($f,e$) need not exist when the one of ($e,f$) exists and that the two distributions need not be identical when they both exist.

The first four of the above identities always define a probability distribution on {0,1}×{0,1} satisfying the fifth identity $p(1,0)+p(1,1)=\mu(e)$, but the last identity $p(0,1)+p(1,1)=\mu(f)$ need not hold. It is equivalent to $e \rightarrow_\mu f$. Therefore, the joint distribution of the event pair ($e,f$) exists if and only if $e \rightarrow_\mu f$ holds.

In the case of relative compatibility $e \leftrightarrow_\mu f$, a second joint distribution $q$ on $\{0,1\}\times\{0,1\}$ exists with $q(1,1)=\mu(f)\mu(e|f)$, $q(1,0)=\mu(f)\mu(e'|f)$, $q(0,1)=\mu(f')\mu(e|f')$, $q(0,0)=\mu(f')\mu(e'|f')$, $q(1,0)+q(1,1)=\mu(f)$, and $q(0,1)+q(1,1)=\mu(e)$. The difference between $p$ and $q$ stems from the sequential order of the two measurements. When the test $e$ versus $e'$ is executed first and the test $f$ versus $f'$ second, the joint distribution is $p$. When the test $f$ versus $f'$ is executed first and the test $e$ versus $e'$ second, the joint distribution is $q$.

Does now $p(k,l)=q(l,k)$ hold for $k,l \in \{0,1\}$ as one would expect from the joint distributions in the case of simultaneous measurability of $e$ and $f$? This leads to the definition of weak comeasurability in the next section.

In the quantum logical approaches, the joint distribution is often defined in a different way by $p(1,1):=\mu(e \wedge f)$, $p(1,0):=\mu(e \wedge f')$, $p(0,1):=\mu(e' \wedge f)$, and $p(0,0):=\mu(e' \wedge f')$. The lattice operation $\wedge$ is used here, requiring the assumption that $E$ is a lattice. Then $p(k,l)=q(l,k)$ always holds for $k,l \in \{0,1\}$, but again $p$ is a reasonable probability distribution only under special conditions. Considering the von Neumann algebras, it becomes immediately evident that the two definitions are different, since $\mu(e)\mu(f|e) = \hat{\mu}(efe)$ and $efe$ is not identical with $e \wedge f$ unless $e$ and $f$ commute.

## 6 Weak comeasurability

In the case of simultaneous measurability of two events $e$ and $f$, it would be expected that their joint distributions exist and satisfy $p(k,l)=q(l,k)$ for $k,l \in \{0,1\}$. Two events $e$ and $f$ shall therefore be called weakly comeasurable in the state $\mu$ if $\mu(a)\mu(b|a)=\mu(b)\mu(a|b)$ holds for all $a,b \in \{e,e',f,f'\}$. They shall now be called weakly comeasurable if they are weakly comeasurable in each state $\mu$. The following lemma shows that the weak comeasurability is stronger than compatibility and identifies a necessary and sufficient condition for the equivalence of weak comeasurability and compatibility.

**Lemma 1:** *The events $e$ and $f$ in a UCP space $E$ are weakly comeasurable in the state $\mu$ if and only if the condition $e \leftrightarrow_\mu f$ and the identity $\mu(e)\mu(f'|e) + \mu(e')\mu(f|e') = \mu(f)\mu(e'|f) + \mu(f')\mu(e|f')$ hold.*

*Proof*: First assume the weak comeasurability. Then $\mu(f|e)\mu(e) + \mu(f|e')\mu(e') = \mu(e|f)\mu(f) + \mu(e'|f)\mu(f) = \mu(f)$ and the same with exchanged roles of $e$ and $f$ such that $e \leftrightarrow_\mu f$. Moreover, $\mu(e)\mu(f'|e) + \mu(e')\mu(f|e') = \mu(f)\mu(e'|f) + \mu(f')\mu(e|f')$ holds since, by the comeasurability, the first summand on the left-hand side coincides with the second one on the right-hand side and the other two summands coincide in the same way.

Now assume $e \leftrightarrow_\mu f$ and $\mu(e)\mu(f'|e)+\mu(e')\mu(f|e') = \mu(f)\mu(e'|f)+\mu(f')\mu(e|f')$. Then $p(1,0)+p(0,1) = q(1,0)+q(0,1)$. Furthermore $p(1,0)+p(0,1)+2p(1,1) = \mu(e)+\mu(f) = q(0,1)+q(1,0)+2q(1,1) = p(0,1)+p(1,0)+2q(1,1)$ such that $p(1,1)=q(1,1)$. Moreover, $p(0,1)+p(1,1) = \mu(f) = q(1,0)+q(1,1)$ such that $p(0,1)=q(1,0)$, and $p(1,0)+p(1,1)=\mu(e)=q(0,1)+q(1,1)$ such that $p(1,0)=q(0,1)$. Finally, $p(0,0) = 1 - p(1,1) - p(1,0) - p(0,1) = 1 - q(1,1) - q(1,0) - q(0,1) = q(0,0)$. □

So the compatibility of the two events $e$ and $f$ in the state $\mu$ implies their weak comeasurability in the state $\mu$ only together with the further condition

$$\mu(e)\mu(f'|e) + \mu(e')\mu(f|e') = \mu(f)\mu(e'|f) + \mu(f')\mu(e|f'). \tag{2}$$

This symmetry condition for the conditional probabilities is well-known and was discovered by Alfsen and Shultz [2] from purely mathematical reasons; they needed its general validity as an additional condition to derive a Jordan algebra structure from their spectral duality. In a similar way it is used in [3] and [4]. A long discussion of condition (2) can be found in [5], where it is shown that the general validity of (2) implies the absence of third order interference as introduced by Sorkin [6]. For the equivalence of compatibility and weak comeasurability, however, it is sufficient that (2) holds for compatible events $e$ and $f$, and its general validity for all event pairs is not necessary.

In the von Neumann algebra setting, $\mu(e)\mu(f'|e)+\mu(e')\mu(f|e') = \hat{\mu}(ef'e+e'fe') = \hat{\mu}(e-efe+f-ef-fe+efe) = \hat{\mu}(e+f-ef-fe)$ and $\mu(f)\mu(e'|f) + \mu(f')\mu(e|f') = \hat{\mu}(e+f-ef-fe)$ in the same way such that (2) always holds, and therefore simultaneous measurability and relative compatibility are equivalent.

## 7 Strong comeasurability

A stronger form of comeasurability is obtained considering the iterated conditional probability $(\mu_e)_f$. The product $\mu(e)\,\mu_e(f)\,(\mu_e)_f(d)$ is the probability that the outcomes in a series of three subsequent measurements are $e$ in the first, $f$ in the second and $d$ in the third measurement, when $\mu$ is the initial state of the system under consideration before the three measurements. Therefore, two events $e$ and $f$ shall be called strongly comeasurable in the state $\mu$ if $\mu(a)\,\mu_a(b)\,(\mu_a)_b(d) = \mu(b)\,\mu_b(a)\,(\mu_b)_a(d)$ holds for all $a,b \in \{e,e',f,f'\}$ and all $d \in E$.

Selecting $d=\mathbb{1}$, it becomes immediately clear that $\mu(a)\mu_a(b)=\mu(b)\mu_b(a)$ for $a,b \in \{e,e',f,f'\}$. The strong comeasurability thus implies the weak one. Moreover, $(\mu_a)_b=(\mu_b)_a$ unless $\mu(a)\mu_a(b) = \mu(b)\mu_b(a) = 0$. This means that two successive measurements with the initial state $\mu$ result in the same final state when the first measurement tests $e$ versus $e'$ and the second one $f$ versus $f'$ or when the first measurement tests $f$ versus $f'$ and the second one $e$ versus $e'$. The sequential order of the two measurements then has no impact on the final state.

The two events $e$ and $f$ shall be called strongly comeasurable if they are strongly comeasurable in each state $\mu$.

## 8 Algebraic compatibility

The last level of compatibility for a pair of events $e$ and $f$ is defined by the existence of three orthogonal events $d_1$, $d_2$, $d_3$ in $E$ such that $e=d_1+d_2$ and $f=d_2+d_3$. This means that $e$ and $f$ lie in a Boolean subalgebra of $E$. The subalgebra consists of sixteen elements: $0$, $\mathbb{1}$, $d_1$, $d_2$, $d_3$, $d_4:=(d_1+d_2+d_3)'$, and the sums of all pairs and triples among $d_1$, $d_2$, $d_3$, $d_4$. In this case, the events $e$ and $f$ shall be called algebraically compatible. Algebraic compatibility implies strong comeasurability. This follows from the following lemma.

**Lemma 2:** *Suppose that $e=d_1+d_2$ and $f=d_2+d_3$ with three orthogonal events $d_1$, $d_2$, $d_3$ in a UCP space $E$. Then* $\mu(e)\mu_e(f)(\mu_e)_f(a) = \mu(d_2)\mu_{d_2}(a) = \mu(f)\mu_f(e)(\mu_f)_e(a)$ *for all events $a$ in $E$ and all states $\mu$ on $E$.*

*Proof:* First suppose $\mu(d_2)=0$. If $\mu(e)=0$, then $\mu(e)\mu_e(f)(\mu_e)_f(a) = 0 = \mu(d_2)\mu_{d_2}(a)$. If $\mu(e)>0$, then $\mu_e(f)=\mu_e(d_2)+\mu_e(d_3)=\mu_e(d_2)=\mu(d_2)/\mu(e)=0$ and again $\mu(e)\mu_e(f)(\mu_e)_f(a) = 0 = \mu(d_2)\mu_{d_2}(a)$. Now suppose $\mu(d_2)>0$. Then

$$\nu := \frac{\mu(e)\mu_e(f)}{\mu(d_2)}(\mu_e)_f$$

is non-negative and coincides with $\mu/\mu(d_2)$ on $\{a \in E : 0 \leq a \leq d_2\}$. Moreover, with $d_4:=(d_1+d_2+d_3)'$, $(\mu_e)_f(d_1)=0$ since $d_1 \perp f$, $(\mu_e)_f(d_2)=\mu(d_2)/(\mu_e(f)\mu(e))$ since $d_2 \perp e'$ and $d_2 \perp f'$, $(\mu_e)_f(d_3)=\mu_e(d_3)/\mu_e(f)=0$ since $d_3 \perp f'$ and $d_3 \perp e$, $(\mu_e)_f(d_4)=0$ since $d_4 \perp f$. Thus $\nu(\mathbb{1})=\nu(d_1)+\nu(d_2)+\nu(d_3)+\nu(d_4)=\nu(d_2)=1$. Therefore $\nu=\mu_{d_2}$. The remaining part follows with exchanged roles of $e$ and $f$. □

## 9  The and-operation

So far, a logical and-combination has been considered only for mutually exclusive events; this is the sum $e+f$ for two orthogonal events $e$ and $f$. Even with the comeasurability of two events, it has not been assumed that a reasonable logical and-combination for the two events exists. If now the event $d$ represented such an and-combination of the events $e$ and $f$, one would expect that $e$ and $f$ are strongly comeasurable and that $\mu(e)\mu_e(f)(\mu_e)_f(a)=\mu(f)\mu_f(e)(\mu_f)_e(a)=\mu(d)\mu_d(a)$ holds for all events $a$ and all states $\mu$.

Under the assumption of Lemma 2, the event $d_2$ satisfies these expectations for a logical and-combination of the events $e$ and $f$. Thus an event "*e and f*" exists for the algebraically compatible event pairs $e,f$. Vice versa, if these expectations are satisfied, are $e$ and $f$ then algebraically compatible? The answer will be given in the next lemma and requires a further condition which more closely couples the state space with the orthogonality relation.

If $e \perp f$ holds with two events $e$ and $f$, then $1=\mu(\mathbb{1}) \geq \mu(e+f)=\mu(e)+\mu(f)$ for all states $\mu$ and particularly $\{\mu \in S : \mu(e)=1\} \subseteq \{\mu \in S : \mu(f)=0\}$. However, in general, $\{\mu \in S : \mu(e)=1\} \subseteq \{\mu \in S : \mu(f)=0\}$ is not sufficient to imply $e \perp f$. This becomes the condition already announced.

**Lemma 3:** *Suppose that $\{\mu \in S : \mu(e)=1\} \subseteq \{\mu \in S : \mu(f)=0\}$ implies $e \perp f$ for events $e$ and $f$ in a UCP space $E$. Then an order relation is defined on $E$ by $e \leq f$ if $e \perp f'$. Moreover, the following three conditions are equivalent for an event pair $e$ and $f$:*
(i)  *The events $e$ and $f$ are weakly comeasurable and there is an event $d$ with $\mu(e)\mu_e(f) = \mu(f)\mu_f(e) = \mu(d)$ for all states $\mu$.*
(ii) *The events $e$ and $f$ are strongly comeasurable and there is an event $d$ with $\mu(e)\mu_e(f)(\mu_e)_f(a) = \mu(f)\mu_f(e)(\mu_f)_e(a) = \mu(d)\mu_d(a)$ for all events $a$ in $E$ and all states $\mu$ on $E$.*
(iii) *The events $e$ and $f$ are algebraically compatible and there are three pairwise orthogonal events $d, d_1, d_2$ such that $e=d_1+d$ and $f=d_2+d$.*

*Proof*: First, it is shown that $\leq$ is an order relation. It is obvious that $e \leq e$ holds for $e \in E$. If $e \leq f$ for $e$ and $f$ in $E$, then $1 \geq \mu(e+f')=\mu(e)+1-\mu(f)$ such that $\mu(e) \leq \mu(f)$ for all states $\mu$. If $e \leq f$ and $f \leq e$, then $\mu(e)=\mu(f)$ for all states $\mu$ and $e=f$ by (UC1). Now suppose $d \leq e$ and $e \leq f$ for three events $d,e,f$. Then $\{\mu \in S : \mu(d)=1\} \subseteq \{\mu \in S : \mu(e)=1\} \subseteq \{\mu \in S : \mu(f)=1\}$, and the condition assumed in Lemma 3 implies that $e \leq f$.

From Lemma 2, it follows that the condition (iii) implies (ii) since, with $d_3:=(d_1+d_2+d)'$, the four events $d_1, d_2, d_3$ and $d$ are pairwise orthogonal and $e=d_1+d$, $f=d_2+d$, $e'=d_2+d_3$ and $f'=d_1+d_3$. Since the strong implies the weak comeasurability, it follows that (ii) implies (i) by selecting $a=\mathbb{1}$.

It remains to show that (i) implies (iii). Suppose that $e$ and $f$ are weakly comeasurable and that there is an event $d$ with $\mu(e)\mu_e(f)=\mu(f)\mu_f(e)=\mu(d)$ for all states $\mu$. Then $\mu(e)=\mu_e(f)=\mu(f)=\mu_f(e)=1$ and $\mu(e')=0=\mu(f')$ for $\mu(d)=1$. The condition assumed in Lemma 3 implies $d \perp e'$ and $d \perp f'$, and by (OS6) there are $d_1$ and $d_2$ which are both orthogonal to $d$ such that $e=d_1+d$ and $f=d_2+d$. Moreover, if $\mu$ is a state with $\mu(d_1)=1$, then $\mu(e)=1$, $\mu(e')=0$, $\mu(d)=0=\mu_e(f)$ and $\mu(d_2)=\mu(f)=\mu_f(e)\mu(f)+\mu_f(e')\mu(f)=\mu_e(f)\mu(e)+\mu_e(f)\mu(e')=0$ such that $d_1 \perp d_2$. □

If $\{\mu\in S : \mu(e)=1\} \subseteq \{\mu\in S : \mu(f)=0\}$ implies $e\perp f$ for the events $e$ and $f$ in a UCP space $E$, then $E$ becomes a partially ordered set where $e\leq f$ holds if and only if $\{\mu\in S : \mu(e)=1\} \subseteq \{\mu\in S : \mu(f)=1\}$. Under these assumptions, two events $e$ and $f$ are weakly or strongly comeasurable and an event $d$ exists which represents a reasonable form of a logical and-combination of $e$ and $f$ if and only if they are algebraically compatible.

## 10 Quantum mechanics

The quantum logic of quantum mechanics with the common Hilbert space formalism is the system of all closed linear subspaces of the Hilbert space or, more generally, the projection lattice in a von Neumann algebra.

Now suppose that $e \rightarrow f$ holds for two elements $e$ and $f$ in the projection lattice of a von Neumann algebra without type $I_2$ part. By equation (1) then $f=efe+e'fe'$ such that $ef=e(efe+e'fe')=efe$ $=(efe+e'fe')e=fe$. This means that $e$ and $f$ commute. Moreover, $(ef)^*=fe=ef$ and $(ef)^2=efef=eeff=ef$ such that $ef$ is a self-adjoint projection. This also holds for $ef'$ and $fe'$. With $d_1:=ef'$, $d_2:=ef$, $d_3:=fe'$, it follows that $e$ and $f$ are algebraically compatible.

Therefore, the weakest one ($e \rightarrow f$) among the five compatibility and comeasurability levels implies the strongest one (algebraic compatibility of $e$ and $f$). Thus all five levels are equivalent and coincide with the common concept of operator commutation. This includes even the first asymmetric level $e \rightarrow f$, which automatically becomes symmetrical in $e$ and $f$ and thus equivalent to $e \leftrightarrow f$. A more general setting where the five compatibility and comeasurability levels still coincide will be presented in the next section.

In the projection lattice of a von Neumann algebra, the infimum $e\wedge f$ exists for each pair of events $e$ and $f$. A reasonable interpretation of $e\wedge f$ as "$e$ and $f$", however, requires the identity $\mu(e\wedge f) = \mu(e)\mu(f|e) = \hat{\mu}(efe)$ for all states $\mu$. This means $e\wedge f=efe$, which holds if and only if $e$ and $f$ commute. Therefore, the infimum $e\wedge f$ should not generally be interpreted as the logical and-combination of the events $e$ and $f$, but only when the two events are compatible, and there is no physically or probabilistically motivated reason to assume from the beginning that a quantum logic should be a lattice.

## 11 P-projections

The same coincidence of the five compatibility and comeasurability levels as in the von Neumann algebras holds for the projection lattice in a so-called JBW algebra without type $I_2$ part. The JBW algebras represent the Jordan analogue of the von Neumann algebras and include exceptional Jordan algebras which cannot be represented as Hilbert space operators [13]. The coincidence of the compatibility and comeasurability levels can be shown for the JBW algebras in a way analogue to the one for the von Neumann algebras or can be concluded from the following lemma where a more general setting is considered. Note that an enhanced version of Gleason's theorem is available also for the JBW algebras [14], [15] and that the projection lattice in a JBW algebra $A$ is identical with the extreme points of the unit interval $[0,\mathbb{1}]:=\{x\in A:0\leq x\leq \mathbb{1}\}$.

A positive linear map $P:A\rightarrow A$ on an order-unit space $A$ with $P^2=P$ is called a P-projection if there is a second positive linear map $P'$ on $A$ with $P'^2=P'$ such that the following four conditions are satisfied:

(a) For $0\leq x\in A$, $Px=0$ holds if and only if $P'x=x$.
(b) For $0\leq x\in A$, $P'x=0$ holds if and only if $Px=x$.
(c) If $\varphi(P\mathbb{1})=\varphi(\mathbb{1})$ holds for a positive linear functional $\varphi$ on $A$, then $\varphi(x)=\varphi(Px)$ for all $x\in A$.
(d) If $\varphi(P'\mathbb{1})=\varphi(\mathbb{1})$ holds for a positive linear functional $\varphi$ on $A$, then $\varphi(x)=\varphi(P'x)$ for all $x\in A$.

The P-projections were introduced by Alfsen and Shultz [16]. Note that $P'\mathbb{1}=\mathbb{1}-P\mathbb{1}$. Examples of P-projections are the maps $Px:=exe$ on a von Neumann algebra or $Px:=2e\circ(e\circ x)-e\circ x$ on a JBW algebra, where $e$ is a projection in both cases (with $e=e^*$ in the first case).

**Lemma 4:** *Suppose that $E=ext[0,\mathbb{1}]$ is the set consisting of the extreme points of the unit interval in an order-unit space $A$ with order unit $\mathbb{1}$ and that, for each $e\in E$, the interval $[0,e]:=\{x\in A:0\leq x\leq e\}$ lies in the closed linear hull of $[0,e]\cap E$. Furthermore, suppose that each state $\mu$ on $E$ can be extended to a positive linear functional $\hat{\mu}$ on $A$ and that there is a P-projection $P_e$ with $P_e(\mathbb{1})=e$ for each $e\in E$.*

*With the orthocomplementation and orthogonality relation defined by $e':=\mathbb{1}-e$ and $e\perp f :\Leftrightarrow e+f\leq \mathbb{1}$ for $e,f\in E$, $E$ then becomes a UCP space; the conditional probability of $f$ under $e$ in the state $\mu$ with $\mu(e)>0$ is $\mu_e(f)=\mu(f|e)=\hat{\mu}(P_ef)/\mu(e)$.*

*Moreover, $e\rightarrow f$ implies the algebraic compatibility for two events $e$ and $f$ in $E$ such that all five compatibility and comeasurability levels coincide.*

*Proof:* First it shall be shown that $e+f\in E$ for $e,f\in E$ with $e+f\leq\mathbb{1}$. Therefore, suppose that $e,f\in E$ with $e+f\leq\mathbb{1}$ and that $e+f=tx+(1-t)y$ with $x,y\in[0,\mathbb{1}]$ and $0<t<1$. Then $e=P_ee\leq P_e(e+f)\leq P_e\mathbb{1}=e$ such that $P_ef=0$ and $e=tP_ex+(1-t)P_ey$. Hence $P_ex=P_ey=e$ such that $P_e(\mathbb{1}-x)=0=P_e(\mathbb{1}-y)$. Thus $\mathbb{1}-x=P_{e'}(\mathbb{1}-x)\leq e'$ and $\mathbb{1}-y=P_{e'}(\mathbb{1}-y)\leq e'$ such that $e\leq x$ and $e\leq y$. Finally, $f=t(x-e)+(1-t)(y-e)$ implies that $x-e=y-e=f$ and $x=y=e+f$.

Next it shall be shown that $e-f\in E$ for $e,f\in E$ with $f\leq e$. Suppose that $e,f\in E$ with $f\leq e$ and that $e-f=tx+(1-t)y$ with $x,y\in[0,\mathbb{1}]$ and $0<t<1$. Then $0\leq P_f(e-f)=P_f(e)-P_f(f)\leq P_f(\mathbb{1})-P_f(f)=f-f=0$ such that $0=P_f(e-f)=tP_fx+(1-t)P_fy$ and hence $0=P_fx=P_fy$. Therefore $x=P_{f'}x\leq f'$ and $y=P_{f'}y\leq f'$. Thus $x+f\leq f'+f=\mathbb{1}$ and $y+f\leq f'+f=\mathbb{1}$. From $t(x+f)+(1-t)(y+f)=e$ it then follows that $x+f=y+f=e$ and $x=y=e-f$.

Suppose now that $e,f,g\in E$ such that $e+f\leq\mathbb{1}$, $e+g\leq\mathbb{1}$ and $f+g\leq\mathbb{1}$. Then $P_g(e+f)=P_ge+P_gf=0$ and $e+f=P_{g'}(e+f)\leq g'$ such that $e+f+g\leq g+g'=\mathbb{1}$.

So far, (OS1) to (OS6) have been shown. Since $A$ is an order-unit space, there are sufficiently many positive linear functionals on $A$ and their restrictions to $E$ yield (UC1). Now (UC2) shall be proven. If $\nu$ is a conditional probability under $e$ in the state $\mu$ with $\mu(e)>0$, then $\nu(e)=1$ and $\nu(f) = \hat{\nu}(P_ef)$ for $f\in E$. Since $0\leq P_ef\leq P_e\mathbb{1}=e$, $P_ef$ lies in the closed linear hull of $[0,e]\cap E$ and therefore $\hat{\nu}(P_ef)=\hat{\mu}(P_ef)/\mu(e)$. That this defines a conditional probability, is immediately clear from the properties of the P-projections.

Finally suppose $e\rightarrow f$. This means $f=P_ef+P_{e'}f$. It shall be shown first that $P_ef\in E$. Assume $P_ef=tx+(1-t)y$ with with $x,y\in[0,\mathbb{1}]$ and $0<t<1$. Then $0=P_{e'}P_ef=\leq tP_{e'}x+(1-t)P_{e'}y$ such that $P_{e'}x=0=P_{e'}y$. Therefore, $x=P_ex\leq e$ and $y=P_ey\leq e$. Hence $0\leq x+P_{e'}f\leq e+e'=\mathbb{1}$ and $0\leq y+P_{e'}f\leq e+e'=\mathbb{1}$. From $f=t(x+P_{e'}f)+(1-t)(y+P_{e'}f)$ it then follows that $x+P_{e'}f=y+P_{e'}f$ and $x=y$. Thus $P_ef\in E$.

Moreover, $P_ef\leq P_e\mathbb{1}=e$ and $P_ef\leq P_ef+P_{e'}f=f$. Now define $d_0:=P_ef$, $d_1:=e-d_0$ and $d_2:=f-d_0=P_{e'}f$. Then $d_1+d_2\leq e+P_{e'}f\leq e+e'=\mathbb{1}$ such that $d_0, d_1, d_2$ are pairwise orthogonal and $e=d_0+d_1$, $f=d_0+d_2$. □

The above lemma covers the projection lattices not only in von Neumann algebras, but also in JBW algebras without type $I_2$ part, including the exceptional ones. In [4], it was shown that each UCP space $E$ can be embedded in the unit interval of an order-unit space similar to the situation of the above lemma, but there are two important differences. In the general case, $E$ need not be identical with the extreme points of the unit interval and the positive projections representing the probability conditionalization need not be P-projections such that the compatibility and comeasurability levels need not coincide.

## 12 Conclusions

The projection lattice in a von Neumann algebra is a special case of a quantum logic with unique conditional probabilities, and here probability conditionalization becomes identical with the state transition of the Lüders - von Neumann measurement process. Therefore, the quantum logics with unique conditional probabilities can be considered an abstract generalized model of the quantum measurement process. This has motivated the introduction of a hierarchy of five compatibility and comeasurability levels. Their meanings are: the absence of quantum interference or influence, simultaneous measurability, and the independence of the final state after two successive measurements from the sequential order of the two measurements. The last and strongest level means that the two events belong to the same Boolean subalgebra of the quantum logic. In the general case, the five compatibility and comeasurability levels seem to differ, but they all coincide in Hilbert space quantum mechanics as well as in the more general setting presented in section 11.

In the common quantum mechanical Hilbert space formalism, the probabilistic interpretation is a later ad-hoc add-on, more or less enforced by the experimental evidence, but not motivated by the mathematical model itself. The quantum logics with unique conditional probabilities involve a clear probabilistic interpretation from the very beginning, which helps to answer questions concerning the theoretical foundations of quantum mechanics. It has been seen in the present paper that this elucidates what a reasonable joint distribution should be and that the lattice operation, if available on a quantum logic, should not generally be interpreted as a logical and-operation.